%Paper: nucl-th/9303009
%From: Avraham Rinat <FNRINAT@WEIZMANN.weizmann.ac.il>
%Date: Sun, 14 Mar 93 18:07:22 +0200

\tolerance 1500
\topskip .75truein
\font\boldm=cmmib10
\font\twelverm =cmr12
\font\twelvei =cmti12
\font\twelveb =cmbx12
\def\Fscr{{\cal F}}
\def\Oscr{{\cal O}}
\def\Nscr{{\cal N}}

\def\kn{{k_0}}

\def\Vscr{{\cal V}}
\def\p3{{d^3p\over {(2\pi)^3}}}
\def\e1{e_0(\vec p)}
\def\e2{e_0(\vec p+\vec q)}

\def\tt{t^{ons}}
\def\zz{\langle g\rangle}
\def\qs{\lower 5pt\hbox to 23pt{\rightarrowfill}\atop
       {s\rarrow\infty}}
\def\qq{\lower 5pt\hbox to 23pt{\rightarrowfill}\atop
      {q\rarrow\infty}}
\def\q|s|{\lower 5pt\hbox to 23pt{\rightarrowfill}\atop
        {|s|\rarrow\infty}}
\def\ton{\lower 5pt\hbox to 23pt{\rightarrowfill}\atop
       {t\rarrow \tt}}
\def\gon{\lower 5pt\hbox to 23pt{\rightarrowfill}\atop
       {g\rarrow \zz}}

\def\bmb{\hbox{\boldm\char'142}}

\def\bmp{\hbox{\boldm\char'160}}
\def\bmq{\hbox{\boldm\char'161}}
\def\bmr{\hbox{\boldm\char'162}}

\def\l{\left (}
\def\r{\right )}
\def\o{\over}
\twelverm
\newdimen\offdimen
\def\offset#1#2{\offdimen #1
   \noindent \hangindent \offdimen
   \hbox to \offdimen{#2\hfil}\ignorespaces}
\parskip 0pt
\rightline{WIS-92/102/Dec PH}
\vskip .6truein

\def\ctrline{\centerline}
\def\({\lbrack}
\def\){\rbrack}
\def\om{\omega}

\def\e{\ell}

\def\pso{\Phi_0}
\def\hq{\hat {\hbox{\boldm\char'161}}}
% This is for reference numbers
\par
\def\tindent#1{\indent\llap{#1}\ignorespaces}
\def\refn{\par\hang\tindent}
\baselineskip 24pt
\centerline{\bf The response of non-relativistic confined systems}
\bigskip
\ctrline {\twelveb S.A. Gurvitz and A.S. Rinat}
\bigskip
\centerline {Weizmann Institute of Science, Rehovot 76100,
Israel}
\vskip 1truein\noindent
\baselineskip 21pt
\par
{\bf Abstract.}

We study the non-relativistic response of a 'diquark' bound by confining
forces, for which perturbation theory in the interaction fails. As
non-perturbative alternatives we consider the Gersch-Rodriguez-Smith
theory and a summation method. We show that, contrary to the case of
singular  repulsive forces, the GRS theory can generally be applied to
confined systems. When expressed in the GRS-West kinematic variable $y$,
the response has a standard asymptotic limit and calculable
dominant corrections of orders $1/q,1/q^2$.
That theory therefore clearly demonstrates
how constituents, confined before and after the absorption of the
transferred momentum and energy, behave as asymptotically free particles.
We compare the GRS results with those for a summation method for harmonic
and square well confinement and also discuss the convergence of the GRS
series for the response in powers of $1/q$.

\vskip.8truein\baselineskip 12pt
%\leftline{\hskip38pt PACS:????}
\vfil
\eject\

\baselineskip 24pt
\par

\par
{\bf 1. Introduction.}
\par
Consider inclusive scattering on a target of mass $M_T$ with confined
constituents. We focus on the response or structure function $S(q,\om)$
and in particular on its limit, when for fixed Bjorken scaling variable
$x=(q^2-\om^2)/2M_T\om$, the momentum-energy transfer $q,\om\to\infty$.
The description of the response in that limit requires a relativistic
theory and one finds for instance in the parton model that the above
limit is that of free constituents.

There remains the  intriguing question how exactly a system,
composed of  confined constituents which can only distribute
the transferred energy to internal  excitations and target recoil but not
to  dissociation, responds  asymptotically  as if  the constituents  were
free.
It is  then tempting to  exploit the simplicity of  non-relativistic (NR)
dynamics,  in the  hope that  it may  illuminate some  features of  this,
intrinsically relativistic problem.

A second incentive to use NR dynamics comes from the
relative ease to describe, for systems bound by regular forces,
the approach of the response to its asymptotic limit.  The situation is
different if those forces, either repulsive or  attractive are singular.
In particular perturbation theory in the interaction  fails and
non-perturbative approaches  have to  be invoked. We already know
that for systems governed by forces which contain a strong short-range
repulsion the  asymptotic limit of  the response  exists, but
differs  from the  same  for  quasi-free constituents$^1$. In
contradistinction, surprisingly
little has been done in the case of singular  attractive, i.e. confining
forces, and those are the main topic of this note.

An example of such a NR  approach is a recent study by Greenberg of the
response of a 'di-quark' bound by a harmonic oscillator potential$^2$. He
found that, in accordance with the naive parton model, the asymptotic
response $qS(q,x)$ in the limit $q\to\infty$, at fixed NR Bjorken
scaling variable $x=q^2/2M\om$ vanishes unless $x$,
which is  the 'quark' momentum  fraction in the infinite  momentum frame,
equals the 'quark'-target mass ratio $m_i/M$. We shall revisit
Greenberg's example in the following.

We start  in Section 2  by scanning NR  descriptions on their  ability to
handle singular forces.  Those theories  routinely employ, instead of the
energy transfer  $\om$, a second  kinematic variable $y$  which differs
from the above
NR Bjorken variable.  Then using the theory of Gersch, Rodriguez
and Smith (GRS)$^3$ we illustrate and emphasize essential
differences in the treatment of singular repulsive, and attractive forces
producing confinement. We show that the GRS theory can handle  the latter
category and  we compute  the response  of 'di-quarks' confined  by an
harmonic oscillator  and by  an infinitely  deep well.   In Section  3 we
generalize a non-perturbative summation technique used by Greenberg $^2$,
compute with it  the same examples and compare the  results.  In addition
we calculate the response for  general forces in a quasi-classical method
and show that the outcome of the  summation method is just the GRS theory
to order $q^{-2}$.   Convergence  conditions  for the  GRS  series  are
discussed  in Section  4.  In  Section 5  we compare  the response,  once
expressed in  terms of the GRS-West  variable $^4$ and then  using the NR
Bjorken  scaling  variable,  and   discuss  the  difference  in  content.
\par
{\bf 2. The GRS series for singular forces.}
\par
We limit ourselves in the following to
'di-quark' targets with constituents of equal mass $m$.
In the target rest system its response per particle, including recoil is
$$S(q,\om)={1\o 2}\sum_n|\Fscr_{0n}(q)|^2
\delta(\om -q^2/4m-E_{n0}),\eqno(1)$$
where
$\Fscr_{0n}(q)=<0|e^{i\vec q\vec r/2}+e^{-i\vec q\vec r/2}|n>$ and
$E_{n0}$ are, respectively, inelastic form
factors and excitation energies. A formal summation over $n$ in
(1) leads to
$$S(q,\om)={1\o{2\pi}}{\rm Im}\sum_{i,j=1,2}<\pso |\exp (-i\bmq\bmr_i)
G(\om)\exp (i\bmq\bmr_j)|\pso>,\eqno(2)$$
with particle and relative coordinates related by $\bmr_{1,2}=\pm\bmr/2$.
The response above contains
$$G(\om)=(\om +E_0-K-V-i\eta)^{-1},\eqno(3)$$
the Greens function of the system in terms of the kinetic energy of
the 'quarks' $K$, the binding energy $E_0$ and the confining interaction
$V$. Regarding the sums in (2)
we recall that the coherent contributions  with $i\ne j$
decrease with increasing $q$ much faster than do incoherent terms with
$i=j$ and the latter will henceforth be disregarded $^4$.

For non-singular regular forces one frequently expands the full
Greens function (3) in a Born series in $V$, thereby using
the free Greens function $G_0=(\om +E_0-K)^{-1}$.
The first term of that expansion is obtained by replacing $G\to G_0$
in Eq. (3) and describes a 'quark', bound before and free after the
transfer of $(q,\om)$.

An alternative approach is due to Gersch, Rodriguez and Smith (GRS), who
showed that the (incoherent part of the) reduced response
$\phi (q,y)\equiv (q/m)S(q,\om )$ for a two-particle target
interacting through local forces can
be written as $^3$
$$\phi (q,y)={1\o{2\pi}}\int_{-\infty}^{\infty}ds
e^{\displaystyle -iys}\int d^3\bmr
\pso (\bmr -s\hq )T_{\sigma}\exp \left\(
i{m\o q}\int_0^s\(K+V(\bmr -
\sigma\hq )-E_0\)d\sigma\right\) \pso (\bmr )d\bmr\eqno(4)$$
Here
$$y=-{q\o 2}+{m\om\o q}\eqno(5)$$
is the non-relativistic GRS-West variable $^{3,4}$, while
$T_{\sigma}$ in Eq. (4) is an operator prescribing $\sigma$-ordering.
Expanding the  exponential in Eq. (4) one obtains
$$\phi (q,y)= F_0(y)+(m/q)F_1(y)+(m/q)^2F_2(y)+\cdots\eqno(6)$$
The first term of the GRS expansion is the asymptotic
limit of  the reduced response
in terms of the single particle momentum distribution $n(p)$
$$F_0(y)={1\o{2\pi}}\int_{-\infty}^{\infty}dse^{\displaystyle -iys}
\int d^3\bmr\pso (\bmr -s\hq )\pso (\bmr)=
2\pi\int_{|y|}^{\infty}n(p)pdp
\eqno(7a)$$
For use below we recall that in the derivation of (7a) one passes the step
$$F_0(y)={q\o m}\int n(p)\delta\left(\om +{\bmp^2\over 2m}-
{(\bmp +\bmq )^2\over 2m}\right) d^3\bmp=2\pi
\int_{|y|}^{\infty}n(p)pdp,\eqno(7b)$$
where the above $\delta$-function describes energy conservation of a
'quark' which before and after the absorption of
($\bmq,\om)$ has the energy of a free, on-shell particle.

For both attractive and repulsive singular interactions, a Born
perturbation theory in $V$ fails. We thus turn to
the GRS series (6), first for singular
{\twelvei repulsive} forces. As an example we choose
an overall, weak binding potential $V(\bmr)$
with a strong, short range repulsion, which for fixed
$\bmb=\bmr_{\perp}, (\hat z=\hat q$) is shown in  Fig. 1 as function of
$z$. For arguments of the wave functions $z-s$ and $z$
on different sides of the hard core, the $\sigma$-integrand in (4)
intersects the hard core region and the corresponding integral
diverges. Consequently for singular repulsion there is no meaning to
the GRS expansion (6). This
does not rule out other non-perturbative approaches,
notably those where a finite $V_{eff}$ replaces the singular $V\,^5$.
One can in fact show that
an asymptotic limit for the response $F_0(y)$ exists, but is not
given by Eq. (7)$^1$.

Also for singular attractive, i.e. {\twelvei confining} interactions
(Fig. 2) the Born series does not exist and we now investigate
whether for those the GRS expression is applicable. Consider first
the wave function arguments $z-s$ and $z$ in (4) for which
$z_1< z-s,z< z_2$, i.e. which lie between the classical turning points
$z_1,z_2$. Then,
although the depth $V_0$ as well as the ground state energy $E_0$
tend to $-\infty$, the difference $V(\bmr -\sigma\hq )-E_0$ remains
finite and so is the $\bmr$ integral in (4). One
reaches the same conclusion if one of the two arguments
above lies outside that region. There $V(\bmr -\sigma\hq )-E_0\to -E_0$
is  unbounded, but one of the wave functions $\pso (\bmr -s\hq )$ or
$\pso (\bmr )$ tends to 0. Since the $\bmr$ integral is finite,
the same is the case with all coefficients $F_n (y)$ in the GRS series
(6). This holds in particular for the asymptotic limit $F_0(y)$ which,
contrary to the case of singular repulsive forces $^1$,
retains the form (7) in terms of the single constituent momentum
distribution $^6$.

The outcome above is surprising since one expects singular attractive and
repulsive forces  to show similar  exceptional behavior (see  Section 5).
We conclude:

i) In contradistinction to the case of repulsive forces, for certain
classes of singular attractive potentials the GRS expansion (6)
for  the response exists and the coefficient functions are finite.

ii) The result for the asymptotic limit of the response $F_0(y)$,
Eq. (7), is the one for free on-shell partons, as if the infinite
potential and binding energy compensate one another.

iii) When for progressively decreasing $q$, increasing distances
are probed the corrections $F_n,\,n\ge 1$ grow in importance.
One may then expect that
qualitatively different behavior sets in only if $q\lambda\approx 1$,
i.e. if the relevant distances in the inclusive scattering become of the
order of a typical length $\lambda$ of $V$. We shall show in
Section 4 that for those values of $q\lambda$ the GRS no more converges.
\par
{\bf 2a. The GRS expansion for selected examples}
\par
We now explicitly demonstrate the applicability of the GRS series for
confining potentials on examples of one-dimensional, two-particle
targets, thereby confirming the above heuristic reasoning.
The general expressions for the first coefficients are $^3$, or
can be transformed to

$$\eqalignno{
F_0(y)&={1\o {2\pi}}\int_{-\infty}^{\infty}dse^{-iys}
\int_{-\infty}^{\infty}dx\pso (x-s)\pso(x)=n(y) & (8a)\cr
\noalign{\vskip5pt}
F_1(y)&={i\o{2\pi}}\int_{-\infty}^{\infty}
dse^{-iys}\int^{\infty}_{-\infty}dx
\pso(x-s)\pso(x)\int^s_0d\sigma\( V(x-\sigma)-V(x)\)
& (8b)\cr
\noalign{\vskip5pt}
F_2(y)&={i^2\o 2\pi}\int_{-\infty}^{\infty}
dse^{-iys}\int^{\infty}_{-\infty}dx
\left\{\pso (x-s)\pso (x){1\o 2}
\left\(\int^s_0d\sigma\( V(x-\sigma)-V(x)\)
\right\)^2\right.\cr
\noalign{\vskip5pt}
&\left. -\left\(\pso''(x-s)\pso (x)-\pso (x-s)\pso''(x)\right\)
\int_0^sd\sigma{s-\sigma\o m}\( V(x-\sigma )-V(x)\)\right\},
& (8c) \cr} $$
with $\pso''=d^2\pso/dx^2$. We now apply the above to harmonic
confinement of the relative motion (Fig. 3). Denoting by $\beta=
(m\om_0/2)^{1/2}$ the inverse length parameter, one finds the following
finite expressions for the first three coefficient functions
(the two lowest order terms had been worked out before $^7$)

$$\eqalign{F_0(y)&={1\o\sqrt{\pi\beta^2}}
\exp (-y^2/\beta^2)\cr
\noalign{\vskip5pt}
(m/q)F_1(y)&=-\l{y\o q}\r\l 1-{2y^2\o 3\beta^2}\r F_0(y)\cr
\noalign{\vskip5pt}
((m/q)^2F_2(y)&=-{1\o 6}\l{\beta\o q}\r^2\l 1-9{y^2\o\beta^2}
+8{y^4\o\beta^4}-{4\o 3}{y^6\o\beta^6}\r F_0(y)\cr}\eqno(9)$$
Next we consider the more intricate case of an infintely deep square well
$V(x)=V_0\theta (a-|x|)$, with $V_0\to -\infty$.
Details are presented in Appendix A and we
present here only the results. With $\gamma(ay)=(\pi/2)^2-(ay)^2$
$$\eqalignno{F_0(y)&={\pi a\over 2}
\left({{\rm cos}(ay)\over{\gamma(ay)}}\right)^2 & (10a)\cr
(m/q)F_1(y)&={2\o qa}\left\(ay-\gamma(ay){\rm tg}(ay)\right\)F_0(y)
&(10b)\cr
%\bye
(m/q)^2F_2(y)&={\pi^2\gamma(ay)\o {4(qa)^2}}\left\(1-{\rm tg}^2(ay)+
{4ay\, {\rm tg}(ay)-1\o{\gamma(ay)}}-{4a^2y^2\o{\gamma^2(ay)}}
\right\) F_0(y) & (10c) \cr}$$
Note the periodic vanishing of $F_0(y), F_1(y)$ for
$ay_n=n\pi/2,n\geq 2$, and for the same $y_n$ the unboundedness
of the ratio $F_1(y)/F_0(y)$ (but not of $F_1(y)$ itself!). Those  are
characteristics of the special properties of the square well potential.
\par
{\bf 3. The non-perturbative summation method}
\par
{\bf 3a. Development}
\par
The non-perturbative expression (1) for the (reduced) response is
in general  quite
impractical, since for regular interactions spectra are predominantly in
the continuum. It is indeed the simplicity of spectra and wave functions
of some confining forces which enables use of, what shall be referred
to as the summation method. The basic assumption to be made in (1) after
use of (5) is
$$\phi(q,y)={q\o {2m}}\sum_{n} |\Fscr_{0n}(q)|^2
\delta\left({yq\o m}+{q^2\o {4m}}-E_{n0}\right)
\rightarrow {q\o {2m}}\left|\Fscr_{o\nu}(q)\right|^2
\Nscr\left(n\to\nu(q,y)\right),\eqno(11)$$
with $\Nscr(n)=\left|dE_{n0}/dn\right|^{-1}$, the level density.
It prescribes  the replacement $n\,\to\nu(q,y)$ everywhere and
subsequent replacement of the sum over discrete $n$ in (1) an integral.
We shall now apply (11) to the cases studied above.
\par
{\bf 3b. Results for selected examples}
\par
For the case of harmonic confinement
$$\eqalign
{\nu(q,y)&=(q^2/2\beta^2)(1+2y/q)\cr
\Nscr(n)&=\om_0}\eqno(12)$$
which leads to
$$\eqalign{\phi(q,y)&={1\o{\sqrt{\pi\beta^2}}}\left\(1-{y\o q}+
{\beta^2\o 2q^2}\left({3y^2\o {\beta^2}}-{1\o 3}\right)+
\Oscr(q^{-3})\right\){\rm exp}\left\(-{q^2\o {2\beta^2}}h(q,y)\right\)\cr
h(q,y)&=-2y/q+(1+2y/q){\rm ln}(1+2y/q)}\eqno(13)$$
For fixed $y$, Eq. (13) allows a large $q$ expansion
$$\phi(q,y)=\(\bar F_0(y)+(m/q)\bar F_1(y)+(m/q)^2\bar F_2(y)
+\Oscr(q^{-3})\),\eqno(14)$$
with $\bar F_i(y)$ coinciding with the GRS coefficients in Eq. (9).

A number of remarks are in order. First, after adjusting constants due
to different definitions of $\phi$, Eqs. (13) and (14)
do not agree with the corresponding result which can be derived from
Eq. (14), Ref. (2).  The difference is due to disregard
there of all but the first two factors in Stirling's formula
$n!=e^{-n}n^n (2\pi n)^{1/2}\big(1-1/12n+\Oscr(n^{-2})\big)$.
Since from (12)  $\sqrt n=(q/\beta)\sqrt{1/2\left\(1+2y/q\right\)}$,
even in the asymptotic limit, $\bar F_0(y)$ should contain the correction
$(2\pi n)^{1/2}$. It incidentally renders $\bar F_0\propto
\beta^{-1}$, the natural length scale for the harmonic
oscillator, and not $\bar F_0\propto \omega_0^{-1}$ as in Ref. (2). The
same also affects the dominant coeffitients $\bar F_1,\bar F_2$: once the
corrections are applied, the lowest three coefficients agree with GRS.

It is of course gratifying to see the correspondence of those
results to $\Oscr (q^{-3})$ by two methods, as different
as the explicit summation in Eq. (11) and the
expression (4). In fact, the agreement should not be taken lightly. On
the one hand it brings to the fore the question of convergence of the
GRS series (see Section 4) and on the other hand  the replacement in
Eq. (9) of a discrete sum over  delta-functions  by an integral.
For regular forces with an overwhelmingly continuous
spectrum, the above replacement seems justified, but this is not
obvious for confining potentials with purely discrete spectra.

Next we turn to the case of a 'di-quark' confined by an infinitly
deep, one-dimensional square well. Eq. (11) yields for that case
$$\phi(q,y)={a^2q\over{2\pi^2 \nu(q,y)}}
\sum_{n\ge 1}\left\(|\Fscr^{(+)}_{0n}(q)|^2
\delta\left(n-{1\o 2}-\nu(q,y)\right)+
|\Fscr^{(-)}_{0n}(q)|^2\delta\left(n-\nu(q,y)\right)\right\),\eqno(15a)$$
where $\gamma(z)=(\pi^2/4)-z^2$ and
$$\pi\nu(q,y)=(aq/2)\left(1+4y/q+(\pi/aq)^2\right)^{1/2}
=(aq/2)\left(1+2y/q+2\gamma(ay)/a^2q^2+\Oscr(q^{-3})\right)\eqno(15b)$$
where
$\Fscr^{(\pm)}_{0n}(q)$ are the inelastic density form factors, linking
the ground state to the excited  even and odd parity
states $a^{-1/2}{\rm cos}\left\(x\pi(n-{1\o 2})/a\right\)$,
respectively $a^{-1/2}{\rm sin}\left\(x\pi n/a\right\)$, for $n\ge 1$.
Proceeding as in Section (3a) one finds after some algebra
$$\left|\Fscr^{(\pm)}_{0n}(q)\right|^2={\pi^2\over 4}
\left|{{\rm cos}(aq/2-\pi\nu(q,y))\over{\left\(aq/2-\pi\nu(q,y)\right\)^2
-\pi^2/4}}\mp
{{\rm cos}(aq/2+\pi\nu(q,y))
\over{\left\(aq/2+\pi\nu(q,y)\right\)^2-\pi^2/4}}\right|^2 \eqno(16)$$
Substituting (16) into (15a) and using (15b), one obtains
$$\eqalignno{
\phi(q,y)&={\pi a\o 2}\left\(1-2y/q+\Oscr(q^{-2})\right\)
\left\(D_1^2(q,y)+D_2^2(q,y)
\right\)& (17a)\cr
D_1(q,y)&={{\rm cos}(ay)\over{\gamma(ay)}}\left\(1+{2y\over q}
\left(1-\gamma(ay)
{{\rm tg}(ay)\over {2ay}}\right)+\Oscr(q^{-2})\right\)& (17b)\cr
D_2(q,y)&={{\rm cos}\(a(q+y+\Oscr(q^{-1}))\)
\over{\pi^2/4-a^2(q+y+\Oscr(q^{-1}))^2}}&(17c)\cr}$$
\noindent
Clearly for fixed $y,\, D_1(q,y)$
permits a large $q$ expansion but $D_2(q,y)\propto {\rm cos}(aq)/q^2$
does not, thus
$$\phi(q,y)=\(\bar F_0(y)+(m/q)\bar F_1(y)+
\Oscr(q^{-2})\)+{a\pi\over 2}D_2^2(q,y)\eqno(18)$$
One then shows that $\bar F_{0,1}(y)=F_{0,1}(y)$ as in Eqs. (10), but
for $n\ge 2,\,\bar F_n(y)\ne F_n(y)$.
This is not surprising since the Euler interpolation formula, the first
term of which gives the replacement of the sum in (11) by an integral, is
only valid for analytic functions. Moreover, for a square well the
density of levels $\Nscr(n)$ grows linearly with $n$, casting doubt on
the appropriate use of the summation method.

{\bf 3c. The quasi-classical response for general $V$.}

We prove in Appendix B the following quasi-classical result for the
reduced response
$$\phi(q,y)=\(F_0(y)+(m/q)F_1(y)+\Oscr(q^{-2})\)\eqno(19)$$
It shows that the application of (11) generallyleads
to the first two terms in the GRS series in the form Eqs. (8a), (8b).
Clearly the above holds only if the quasi-classical method is at
all applicable. This is for instance not the case for the square well
treated in the previous section. When nevertheless worked out for
that case, a non-analytic term like $D_2$ in (17c) appears also in this
treatment.

The above result brings to mind Rosenfelder's  treatment of the response
using Wigner distribution functions $^8$. It had been  observed before
that the approximation which Rosenfelder suggested and which uses
another aspect of the semi-classical approach $^9$, also
produces $F_0$ and the  correct dominant correction
$F_1(y)$, but not higher order coefficients.

\par
{\bf 4. Convergence of series expansions fo the response}
\par
Little is known on the convergence of various series expansions
for the response. We mention a proof that the reduced response, when
expressed in an alternative $'$plane
wave$'$ kinematic variable $y_0$ instead of $y$, Eq. (5),
converges to the plane wave impulse limit $\phi^{PWIA}(q,y_0)$ (and
in fact to the asymptotic limit $F_0(y_0)$, Eq. (7)) provided the
interaction has finite norm $||V||\,^{10}$. This sufficient condition
does not distinguish between attractive and repulsive forces and excludes
singular $V$ of either type.

In the light of the above stands the remarkable observation above that
for classes of
confining forces, the exponent in the GRS expression (4) for the
response exists. Again this is a necessary but not a sufficient
condition for the convergence of the $1/q$ expansion (6). No doubt
that for each system there are additional conditions which depend
on dimensionless quantities. Those  can be constructed from the external
momenta $y,q$ and lengths $\lambda$ in the interaction $V$.

In fact the two examples treated are illuminating. First, for both one
observes that $\phi_n(q,y)\equiv (m/q)^nF_n(y)$ is independent of $m$.
It had been remarked before, that although
naturally appearing in the GRS theory $^3$, the ratio $m/q$ cannot be
an expansion parameter$^{11}$: As the above results (9) and (10) show,
the explicit mass of the constituents appears to cancel out in
$\phi(q,y)$, but it may well be implicit in length parameters like
$\lambda=(m\om_0)^{-1/2}$ for the harmonic oscillator.

We now focus on  $F_1,F_2$  in  Eqs. (9)  and (10) which dominate the
approach to the asymptotic limit $F_0$ for non-vanishing,
not too large $y$.
We concentrate on $y=0$ for which $F_1(0)=0$ and the convergence is
fastest.  For the two examples considered, one has

$$(m/q)^2F_2(0)/F_0(0)=\cases{-(\beta^2/6q^2), & for HO\cr
\noalign{\vskip5pt}
\pi^2(\pi^2-4)/16q^2a^2, & for SqW\cr} \eqno(20)$$
The right hand side gives the size of corrections to the latter,
governed by $q\lambda$. The condition $q\lambda>>1$
coincides with the condition, already mentioned in the paragraph before
Section 2b. For small, finite $y$ one has to add $y/q<<1$.

\par
{\bf 5. Response of confined systems in terms of
the Bjorken variable.}
\par

Until here we studied the reduced response expressed in terms of the NR
GRS-West variable (5). In his treatment of NR harmonic confinement
Greenberg used instead a NR Bjorken scaling variable
$$x=q^2/4m\om\eqno(21)$$
with
$$y=(q/4)(x^{-1}-2)\eqno (21')$$
giving the relation to the GRS-West variable. One then shows
that for harmonic confinement the summation method produces for large $q$
$$\tilde\phi(q,x)\equiv (q/m)S(q,\om)
=z(x,q)e^{-(q/q_0)(x^{-1}-2)^2},\eqno(22)$$
with $z$ a regular function of $1/q$. The corresponding reduced response,
when expressed  as function of $x$, has a vanishing asymptotic limit,
unless $x=1/2,\,^2$. The latter is for the equal mass case
the momentum fraction of the 'quarks' in the Galilei-boosted, infinite
momentum frame. We now show that the same conditioned
vanishing in fact holds for any interaction, regular or confining.

Using (21$'$), the asymptotic limit (7b) for $q\to\infty$
can as follows be expressed in $x$
$$qF_0\(y(q,\om)\)=q\tilde F(q,x)=4\int d\bmp \,n(p)\delta\left(
x^{-1}-2-{4p_z\o q}\right)\to\delta\left(x-{1\o 2}\right), \eqno(23)$$
where $n(p)$ drops out due to its normalization. Therefore,
starting from (7b) which is valid for the above forces, is the response
for free, on-shell particles, one reaches the last
identity, namely  an asymptotic response with zero support, except for
$x=1/2,\,^4$. We now ask whether the converse is also true. By way of
example we take a constituent which before absorption of $q$ is off-shell
with energy $e(p)=p^2/2m+\Vscr(p)$: $q\tilde F(q,x)$
has for $q\to\infty$ the $same$ asymptotic limit $\delta(x-1/2)$. One
thus concludes that the response in the $x$ variable  of the form
(23) is no evidence of asymptotically free, on-shell particles,
whereas this is the case for the same in the $y$ variable (7).

The poor content of the asymptotic limit of the response
in terms of the NR Bjorken
variable (21) contrasts with the same in $y$, Eq. (7), which as
function of $y$ enables the extraction of the momentum
distribution of the constituents and the study of the approach towards
that limit. In contradistinction, Eq. (22), expressed in
the NR Bjorken variable, does not permit a series expansion in $1/q$.
Therefore, no matter what $V$ is, the use of $y$
is preferable over the NR Bjorken variable $x\,\,\,^4$.

We close with a remark on the limited, singular
support of the asymptotic limit of the reduced response in the NR Bjorken
variable. Clearly any $p$-dependent term in the $\delta$ argument in (23)
which does not vanish for asymptotic $q$, produces a finite support. For
instance, using relativistic kinematics ($e_p=\sqrt{p^2+m^2}$)
in (7b) as well as the relativistic Bjorken variable
$x_r=(q^2-\om^2)/(4m\om)$  produces a proper $x$-support.$^4$
\par
{\bf 6. Summary}

We discussed  above the  non-relativistic reponse or
structure function  of two-body
systems  of  confined  constituents.    Due  to  their  singular  nature,
perturbation theory in $V$ fails  and non-perturbative methods are called
for. We have investigated the GRS theory, which leads to a
formally exact series expansion for the reduced response in powers of
$1/q$. We could demonstrate that, contrary to the case of singular
repulsive forces, for a system with singular confining forces that theory
may make sense. As a consequence it permits, for  fixed GRS-West variable
$y$, a  power series expansion in  $1/q$. In particular
the  asymptotic limit  is shown to  be the  one for
free,   on-shell  constituents:   For  the   smallest  distances  probed
the response is just not sensitive to confinement of finite range. Higher
order coefficients, relevant for increasing probed distances, correct the
asymptotic limit as if the basic forces were regular.

A second method utilizes the occasional simplicity of spectra  and  wave
functions  of   confined  systems and calculates the response in the form
Eq. (1) by an explicit summation over intermediate excited states.
We then compared the  outcome for  the response in  the two  methods for
examples  of targets  with confined constituents. In  addition we  showed
that, whenever  applicable, the  semi-classical response agrees  with the
GRS series to $\Oscr(q^{-2})$.

Next we tested the GRS series on its convergence in particular for $y=0$.
Convergence conditions require $y$ to be small compared to typical
inverse lenghts in $V$. Finally
we compared the above responses if the GRS-West kinematic variable $y$
is replaced by the non-relativistic Bjorken scaling variable $x$. The
asymptotic limit vanishes except for
values of $x$, equal to the momentum fraction of the constituents in
the Galilean-boosted, infinite momentum frame. No additional information
is contained in that limit, in contradistinction to the one
in the GRS-West variable which contains the single constituent momentum
distribution. For non-relativistic dynamics the above clearly favors
the use of the GRS-West variable over the NR Bjorken variable.

Our concluding remark regards a conjecture of Greenberg, holding that
a Gaussian decrease with $q$ of $\phi(q,x)$ around $x=1/2$ reflects the
rapid vanishing of the $'$quark-quark$'$ interaction for decreasing
separation $^2$. It is instructive to transcribe the above behavior,
using $y$ instead of $x$. Thus $\tilde \phi(q,x)\to\phi(q,y)
\propto {\rm exp}\(-(y/\beta)^2\)$, i.e. a Gaussian in $y$. We now claim
that such behavior need not at all be related to inter-constituent forces
vanishing with $r$: As an example we consider
liquid $^4$He with overall weak, attractive inter-atomic force with a
very strong, short-range repulsion. The resulting single particle
momentum distribution close to $T=0^{\circ}$ is roughly Gaussian $^{12}$
and so is the asymptotic response $F_0(y)$ as is also the case for
harmonic confining forces (cf. Eq. (8a)).

\noindent

{\bf Acknowledgement}

The authors thank M. Kugler for enlightening remarks on the subject
matter.
\par
{\bf Appendix}
\par
{\bf A. Terms in the GRS series for an infinitly deep square well}
\par
The ground state wave function for a square well potential
$V(x)=V_0\theta(a-|x|)$ is
$$\pso (x)=\cases{(1/\sqrt{a})\cos (\kn x), &for $|x|\leq a$\cr
\noalign{\vskip5pt}
(\kn /\sqrt{ma|V_0|)}\exp\left\(
-\sqrt{m|V_0|}(|x|-a)\right\), &for $|x|>a$\cr}\eqno(A1)$$
where $\kn =\sqrt{m(E_0-V_0)}\to\pi/2a$ for $V_0\to -\infty$.
Substituting the above $\pso$ in Eq. (8a) one finds in the limit
$V_0\to\infty$
$$F_0(y)={\pi a\o2}\left\({{\rm cos}(ay)\o{\gamma(ay)}}\right\)^2,
\eqno(A2)$$
where $\gamma(ay)=\pi^2/4-a^2y^2$.
All higher order coefficients contain the singular potential. Notice
that in the limit $V_0\to -\infty$, the integrands in
Eqs. (8b), (8c) vanish for $|s|>2a$, since the product of both wave
functions decreases there  exponentially with $|V_0|$. In general
$F_n(y)$ draws only on that interval
$$F_n(y)={i^n\o 2\pi}\int_{-2a}^{2a}dse^{-iys}R_n(s),\eqno(A3)$$
where $R_n(s)$ denote $x$-integrals (cf. Eqs. (8b), (8c)). Consider first
the $x$-integration in Eq. (8b) over the interval $0\leq s\leq
2a$. For $-a+s\leq x\leq a$ the difference $V(x-\sigma )-V(x)=0$ and the
integral over the remaining $x$-sections can be written as
$$\eqalign{
R_1(s)=&\left\(\int_{-\infty}^{-a+s}dx+\int_a^{\infty}dx\right\)
\pso (x-s)\pso (x)\int^s_0d\sigma\( V(x-\sigma)-V(x)\)\cr
\noalign{\vskip5pt}
=&V_0\int_{-\infty}^{-a+s}\pso (x-s)\pso (x)(x+a-s)dx+
V_0\int_a^{\infty}\pso (x-s)\pso (x)(s+a-x)dx\cr}\eqno(A4)$$
Integrating by parts, one finds that only the second integral contributes
in the limit $V_0\to -\infty$
$$\lim_{V_0\to -\infty}R_1(s)= -{\kn s\o ma}\sin (\kn s)\eqno(A5)$$
Likewise for the second interval $-2a\leq s\leq 0$
$$\lim_{V_0\to -\infty}R_1(s)= {\kn s\o ma}\sin (\kn s)\eqno(A6)$$
Substitution of Eqs. (A5), (A6) into Eq. (A3) produces
$$(m/q)F_1(y)={1\o {qa}}\left\(ay-\gamma(ay){\rm tg}(ay)\right\)
F_0(y)\eqno(A7)$$
In a similar way one computes the third GRS coefficient function
$F_2(y)$, Eq. (8c). For $0\leq s\leq 2a$ only the interval $x>a$
contributes to $R_2(s)$ (cf. Eq. (A4))
$$\eqalign{
R_2(s)=&V_0^2\int_a^{\infty}\pso (x-s)\pso (x){(s+a-x)^2\o 2}dx\cr
\noalign{\vskip5pt}
-&V_0\int_a^{\infty}\(\pso'' (x-s)\pso (x)-\pso (x-s)\pso'' (x)\)
{(s+a-x)^2\o 2m}dx\cr}\eqno(A8)$$
Since for $x\geq a$ one has $\pso''(x)=-mV_0\pso (x)$, the first integral
in Eq. (A8) proportional to $V_0^2$ cancels against the last term in the
second integral. Consequently only the first term in the second integral
survives. Integration by parts gives
$$\lim_{V_0\to -\infty}R_2(s)=-{\kn^3s^2\o 2m^2a}\sin (\kn s)\eqno(A9)$$
As was the case for $R_1$ above (cf. Eqs. (A5) and (A6)),
the region $-2a\leq s\leq 0$ produces (A9) but with the opposite sign.
Substituting $R_2(s)$ into Eq. (A3) yields
$$(m/q)^2F_2(y)={\pi^3\gamma(ay)\o {(4qa)^2}}\left\(1-{\rm tg}^2(ay)+
{4ay\, {\rm tg}(ay)-1\o{\gamma(ay)}}-{4a^2y^2\o {\gamma^2(ay)}}
\right\)
F_0(y)\eqno(A10)$$
Eqs. (A2), (A7) and (A10) are the results cited in Eq. (10).
\par
{\bf B. The quasi-classical response.}
\par

Neglecting wave functions of excited states in the classically forbidden
region, we have inside the classical turning points $x_1,x_2$
$$\eqalign{
\Phi_n(x)&\approx \sqrt {m\o {\pi p_n(x)\Nscr(n)}}{\rm cos}
\left\(\int_{x_1}^x d\xi p_n(\xi)-{\pi\o 4}\right\)\cr
p_n(x)&=\sqrt{m(E_n-V(x))}
\to {q\o 2}\left\(1+{2y\o q}+{2(mE_0-y^2)\o {q^2}}
-{2mV(x)\o{q^2}}+\Oscr(q^{-3})\right\),}\eqno(B1)$$
with $\Nscr$ as in Eq. (11) the density of states. In line with the
summation method, we  used above the $\delta$ function in (1) and the
definition  (5) of $y$. Aiming at a calculation to
$\Oscr (q^{-2})$, one finds for the reduced response (11)
$$\eqalign{
\phi (q,y)&=
{2\o \pi}\left( 1-{2y\o q}\right)
\left|\int_{x_1}^{x_2}dx\pso (x) e^{i{qx\o 2}}
{1\o 2}\left\{\exp\left\(
-i{q\o 2}\left( 1+{2y\o q}+{2mE_0-2y^2\o q^2}\right)
(x-x_1)\right.\right.\right.\cr
\noalign{\vskip5pt}
&\left.\left.\left.
+i{m\o q}\int^x_{x_1}d\xi V(\xi )+i{\pi\o 4}\right\)
+{\rm c.c.}+\Oscr (q^{-2})\right\}\right|^2,\cr}\eqno(B2)$$
where  the density  of states cancels.   Consider first
contributions which  come from  the second ($'$c.c.$'$) term
in the  above bracket. It is
readily seen  that those contribute  to $\phi$ terms proportional  to the
elastic  form factor  or its  square. The former decreases normally as
$1/q^2$ and  can be neglected in  comparison with the first  term in the
brackets in (B2) $^{13}$. To the desired order
$$\eqalign{
\phi (q,y)&={1\o 2\pi}\int\int dx\,dx'\,\pso (x)\pso (x')
\left\( 1-{2y\o q}-i{m\o q}(x-x')\left( E_0-{y^2\o m}\right)\right.\cr
\noalign{\vskip5pt}
&\left. +i{m\o q}\int^x_{x'}d\xi V(\xi )+
 \Oscr (q^{-2})\right\) e^{iy(x'-x)}\cr}\eqno(B3)$$
Using $-y^2/m=(1/m)(d^2/dx^2)e^{-iyx}$ and the Schroedinger equation,
and integrating by parts one finds
$$\phi(q,y)={1\o {2\pi}}\int\int dx\,dx'\,\pso (x)\pso(x')
\left\(1+i{m\o q}\int^{x}_{x'}d\xi V(\xi)-i{m\o q}(x-x')V(x)
+\Oscr(q^{-2})\right\)e^{iy(x-x')}
\eqno(B4)$$
Writing $s=x-x'$ and replacing $\xi\to x-\sigma$ one shows that (B4)
is Eq. (19) with $F_{0,1}(y)$ as in Eqs. (8).

\par

%\noindent
%\vfil
%\eject

{\bf References}
\bigskip

\refn{$^1$}
J.J. Weinstein and J.W. Negele, Phys. Rev. Lett. {\bf 49} 1016 (1982);
S.A. Gurvitz, A.S. Rinat and R. Rosenfelder, Phys Rev. {\bf C40} 1361
(1989); J. Besprosvany and S.A. Gurvitz,  preprint WIS 92/31.

\refn{$^2$}
O.W. Greenberg, Phys. Rev. {\bf D47}, to be published.

\refn{$^3$}
H.A. Gersch, L.J. Rodriguez and Phil N. Smith, Phys. Rev. {\bf A5} 1547
(1972); H.A. Gersch and  L.J. Rodriguez, Phys. Rev. {\bf A8} 905 (1973).

\refn{$^4$}
G.B. West, Phys. Rep. {\bf C18} 264 (1975); Proceedings of International
School on Nuclear Dynamics, Dronten, Netherlands (1988).

\refn{$^{5}$}
R.N. Silver, Phys. Rev. {\bf B37} 3794 (1988); $ibid$ {\bf B38} 2283
(1988); A.S. Rinat $ibid$ {\bf B40} 6625 (1989); A.S. Rinat and
W.H. Dickhoff, $ibid$ {\bf B42} 10004 (1990).

\refn{$^{6}$}
We are of course aware of the  heuristic nature of the above reasoning,
where statements are made  about the finite product of vanishing small
and  unbounded quantities. In an application in Section 2a we shall
rigorously derive those statements.

\refn{$^{7}$}
A.S. Rinat and R. Rosenfelder, Phys. Lett. {\bf B193} 411 (1987).

\refn{$^{8}$}
R. Rosenfelder, Nucl. Phys. {\bf A459} 452 (1986).

\refn{$^9$}
A.S. Rinat, Phys. Rev. {\bf B36} 5171 (1987)

\refn{$^{10}$}
E. Pace, G. Salm$\grave{\rm e}$ and G.B. West,
Phys. Lett {\bf B273} 205 (1991).

\refn{$^{11}$}
A.S. Rinat and M.F. Taragin, Phys. Rev. {\bf B41} 4247 (1990).

\refn{$^{12}$}
P.A. Whitlock and R.M. Panoff, Can. J. Phys. {\bf 65} 1409 (1987);
D.P. Ceperley and E.L. Pollock, $ibid$ {\bf 65} 1416 (1987).

\refn{$^{13}$}
The square well is a notable exception and leads to slower decreasing and
oscillating contributions like $D_2^2$ in (18).

\bigskip
{\bf Figure captions.}

Fig. 1. Cut of a weakly binding potential $V(\bmb,z)$
with strong short range repulsion as function of $z$ for fixed $b$.

Fig. 2. Same as Fig. 1 for a confining $V(\bmb,z)$.

Fig. 3. Harmonic confining potential
energy $E_0,\,\,V$ is defined as the $V_0\to -\infty$ limit of $V(x)=
m\om_0^2x^2/4+V_0$ for $|x|<2\left(|V_0|/m\om_0^2\right)^{1/2}$
and $V=0$ for $|x|\ge 2\left(|V_0|/m\om_0^2\right)^{1/2}$ and
$E_0=V_0+\om_0/2$. $V_0$  drops out of the expressions (8).

\end